\documentclass[aps,prd,onecolumn,groupedaddress,showpacs,nofootinbib,amssymb]{revtex4-2}
\usepackage[dvips]{graphicx}
\usepackage{amssymb}
\usepackage{amsmath}
\usepackage{graphicx,,color}
\usepackage{amsfonts}
\usepackage{bm}
\usepackage{cancel}
\usepackage{comment}
\usepackage{floatflt}

\newcommand\be{\begin{equation}}
\newcommand\ee{\end{equation}}

\allowdisplaybreaks[4]

\begin{document}

\tolerance=5000

\title{Kinetic Axion $F(R)$ Gravity Inflation}
\author{V.K. Oikonomou,$^{1}$}
\affiliation{$^{1)}$ Department of Physics, Aristotle University
of Thessaloniki, Thessaloniki 54124, Greece}
\email{voikonomou@auth.gr;v.k.oikonomou1979@gmail.com}


 \tolerance=5000

\begin{abstract}
In this work we investigate the quantitative effects of the
misalignment kinetic axion on $R^2$ inflation. Due to the fact
that the kinetic axion possesses a large kinetic energy which
dominates its potential energy, during inflation its energy
density redshifts as stiff matter fluid and evolves in a
constant-roll way, making the second slow-roll index to be
non-trivial. At the equations of motion level, the $R^2$ term
dominates the evolution, thus the next possible effect of the
axion could be found at the cosmological perturbations level, via
the second slow-roll index which is non-trivial. As we show, the
latter elegantly cancels from the observational indices, however,
the kinetic axion extends the duration of the inflationary era to
an extent that it may cause a 15$\%$ decrease in the
tensor-to-scalar ratio of the vacuum $R^2$ model. This occurs
because as the $R^2$ model approaches its unstable quasi-de Sitter
attractor in the phase space of $F(R)$ gravity due to the $\langle
R^2 \rangle $ fluctuations, the kinetic axion dominates over the
$R^2$ inflation and in effect the background equation of state is
described by a stiff era, or equivalently a kination era,
different from the ordinary radiation domination era. This in turn
affects the duration of the inflationary era, increasing the
$e$-foldings number up to $5$ $e$-foldings in some cases,
depending on the reheating temperature, which in turn has a
significant quantitative effect on the observational indices of
inflation and especially on the tensor-to-scalar ratio.
\end{abstract}


\maketitle

\section{Introduction}

Dark matter, along with inflation, dark energy and the mysterious
reheating-early radiation domination era, are the mysteries of
modern theoretical physics. These problems have been puzzling
theoretical physicists for decades and to date no definite answer
is given to answer the questions imposed by these problems. Of the
above evolution eras of our Universe, only the dark energy era has
been observationally verified, whereas the rest of the eras remain
still at the speculation level. However, inflation and the closely
related post-inflationary reheating era, are going to be severely
scrutinized in the next fifteen years, by both stage-4 Cosmic
Microwave Background (CMB) experiments
\cite{CMB-S4:2016ple,SimonsObservatory:2019qwx} and by
interferometric and not only gravitational waves experiments like
the LISA, DECIGO, BBO, Einstein telescope and so on
\cite{Hild:2010id,Baker:2019nia,Smith:2019wny,Crowder:2005nr,Smith:2016jqs,Seto:2001qf,Kawamura:2020pcg,Bull:2018lat},
see also \cite{LISACosmologyWorkingGroup:2022jok}. The
gravitational wave interferometric experiments will directly probe
tensor modes which reentered the Hubble horizon during the
mysterious reheating era, thus small wavelength modes that carry
information for both the observational indices of inflation and
also for the reheating era, while the stage-4 CMB experiments will
seek for the $B$-mode polarization in the CMB. The B-modes can be
generated by two distinct effects, by the $E$-mode conversion to
$B$-modes via gravitational lensing for small angular scales or
large-$\ell$ CMB modes, or by tensor modes for large angular
scales or small-$\ell$ CMB  modes. Dark matter though seems
unreachable to us for the time being. Although there are many
proposals for dark matter
\cite{Bertone:2004pz,Bergstrom:2000pn,Mambrini:2015sia,Profumo:2013yn,Hooper:2007qk,Oikonomou:2006mh},
currently it  still remains a mystery what dark matter is
comprised of. The basic known facts about dark matter is that it
has a particle nature, based on observations like the bullet
cluster, and the dark matter particle definitely has a small mass.
An appealing candidate, as elusive among other particles as dark
matter itself, is the axion
\cite{Preskill:1982cy,Abbott:1982af,Dine:1982ah,Marsh:2015xka,Sikivie:2006ni,Raffelt:2006cw,Linde:1991km,Co:2019jts,Co:2020dya,Barman:2021rdr,Marsh:2017yvc,Odintsov:2019mlf,Nojiri:2019nar,Nojiri:2019riz,Odintsov:2019evb,Cicoli:2019ulk,Fukunaga:2019unq,Caputo:2019joi,maxim,Chang:2018rso,Irastorza:2018dyq,Anastassopoulos:2017ftl,Sikivie:2014lha,Sikivie:2010bq,Sikivie:2009qn,Caputo:2019tms,Masaki:2019ggg,Soda:2017sce,Soda:2017dsu,Aoki:2017ehb,Masaki:2017aea,Aoki:2016kwl,Obata:2016xcr,Aoki:2016mtn,Ikeda:2019fvj,Arvanitaki:2019rax,Arvanitaki:2016qwi,Arvanitaki:2014wva,Arvanitaki:2014dfa,Sen:2018cjt,Cardoso:2018tly,Rosa:2017ury,Yoshino:2013ofa,Machado:2019xuc,Korochkin:2019qpe,Chou:2019enw,Chang:2019tvx,Crisosto:2019fcj,Choi:2019jwx,Kavic:2019cgk,Blas:2019qqp,Guerra:2019srj,Tenkanen:2019xzn,Huang:2019rmc,Croon:2019iuh,Day:2019bbh,Odintsov:2020iui,Nojiri:2020pqr,Odintsov:2020nwm,Oikonomou:2020qah}.
With the terminology axion, we do not refer to the QCD axion, but
to an axion like particle, in which case the primordial $U(1)$
Peccei-Quinn symmetry of the axion is broken during inflation, and
the axion develops a non-zero vacuum expectation value. The axion
is a light scalar field, thus it is highly motivated from a string
theory point of view, since scalar fields are the string moduli,
which are a basic and profound characteristic of string theory.
The axion is an elusive particle with extremely small mass, and
admittedly quite hard to detect, however there are direct and
indirect ways to detect it, for example from observations of
neutron stars \cite{Kavic:2019cgk}, or due to black hole
superradiance effects
\cite{Arvanitaki:2019rax,Arvanitaki:2016qwi,Arvanitaki:2014wva,Arvanitaki:2014dfa}.
However, in the future it might also be detected on ground
experiments, utilizing the conversion of the axion to photons in
the presence of a strong magnetic field. The most fascinating
feature of the axion and of axion like particles in general is the
fact that during inflation, the primordial $U(1)$ symmetry is
broken, thus allowing the axion to have a large vacuum expectation
value, and no cosmic string remnants pollute the post-inflationary
era. Another fascinating fact about the axion is that when the
Hubble rate of the Universe becomes of the same order as the axion
mass, the axion commences oscillations around its vacuum
expectation value and its energy density redshifts as $\rho_a\sim
a^{-3}$ thus as cold dark matter. Hence the axion can be the
predominant component of cold dark matter in the Universe.

Modified gravity
\cite{reviews1,reviews2,reviews3,reviews4,reviews5} offers an
appealing theoretical framework, in the context of which the dark
energy era and the inflationary era can be described in a unified
way, see the pioneer article \cite{Nojiri:2003ft} for the first
attempt toward this direction and also Refs.
\cite{Nojiri:2007as,Nojiri:2007cq,Cognola:2007zu,Nojiri:2006gh,Appleby:2007vb,Elizalde:2010ts,Odintsov:2020nwm,Sa:2020fvn}
for later developments in this research line. In both the
inflationary era and dark energy era general relativistic
descriptions, the use of a scalar field, minimally or
non-minimally coupled to gravity, is inevitably needed. With
regard to the inflationary era, the scalar field description can
be somewhat problematic, since the scalar field must inevitably be
coupled to the Standard Model particles, and the couplings are
arbitrary. Also with regard to the dark energy era, the scalar
field description is problematic, since the dark energy equation
of state (EoS) parameter is allowed to take values beyond the
phantom divide line, so it can be less than -1. The scalar field
description of such an evolution requires tachyon fields, which
are not appealing at all in any context. Modified gravity in its
various forms offers a consistent framework which can describe
both the inflationary and the dark energy era, without the
shortcomings of the scalar field description.

Among all the modified gravities, the $f(R,\phi)$ theories are the
most motivated, since in a fundamental primordial scalar field in
its vacuum configuration, the first quantum corrections are higher
powers of the Ricci scalar, see \cite{Oikonomou:2022bqb} for more
details, also combinations of the Ricci scalar with Riemann and
Ricci tensors, such as the Einstein-Gauss-Bonnet theories. In this
article we shall assume that the inflationary era is controlled by
an $F(R)$ gravity in the presence of a primordial axion field. The
axion field shall be assumed to be the misalignment axion, in
which case the primordial $U(1)$ Peccei-Quinn symmetry that the
axion possessed is broken during inflation. There are two
misalignment axion models in the literature, the canonical
misalignment axion \cite{Marsh:2015xka} and the kinetic
misalignment axion models
\cite{Co:2019jts,Co:2020dya,Barman:2021rdr}. The difference
between the two models is that during inflation, the canonical
misalignment axion possesses no kinetic energy, and on the
contrary in the case of the kinetic misalignment axion case, the
axion possesses a large kinetic energy, which dominates the
potential energy. Thus the axion oscillations in the latter case
commences much more later, compared to the former axion model. In
this work we shall investigate the effects of the kinetic
misalignment axion model on the inflationary era generated by an
$F(R)$ gravity and specifically on the $R^2$ inflationary era. At
the equations of motion level, the effects are absent, however at
the cosmological perturbations level, the axion may affect
directly the evolution via the second slow-roll parameter
$\epsilon_2$. As we show, the kinetic axion obeys a constant-roll
evolution, which dominates the evolution at the late stages of the
$R^2$ controlled inflationary era. The kination era caused by the
kinetic axion basically dominates over the $\langle R^2\rangle $
fluctuations which destabilize the inflationary quasi-de Sitter
vacuum. This causes the total EoS of the Universe to be described
by a short kination era, described by a stiff perfect fluid
evolution, which eventually affects the total number of the
$e$-foldings. Remarkably though, the fact that the scalar field
obeys a constant-roll evolution during inflation, does not affect
at all the observational indices of inflation, since the
contribution of the slow-roll parameter $\epsilon_2$ is elegantly
cancelled.

This paper is organized as follows: In section II we present the
essential features of the kinetic axion $F(R)$ gravity model. We
describe in brief the kinetic axion model, and we also present the
way in which the axion post-inflationary may mimic cold dark
matter. In section III, we investigate in detail the inflationary
dynamics of the kinetic axion $F(R)$ gravity model. We show how
the constant-roll evolution of the kinetic axion during inflation
eventually leaves unaffected the dynamics of inflation at the
cosmological perturbations level, and also we show how the
kination era at the last stages of the $R^2$ controlled
inflationary era, eventually prolongs the inflationary era,
increasing the total number of $e$-foldings. The conclusions of
this work follow in the end of the article.

\section{Essential Features of the $F(R)$ Gravity-Kinetic Axion Model}

Before we get to the study of the inflationary dynamics for the
$F(R)$ gravity-kinetic axion model let us first present the
theoretical framework of the model in some detail. The $F(R)$
gravity-kinetic axion model is basically an $f(R,\phi)$ gravity
theory, in which case the gravitational action has the following
form,
\begin{equation}
\label{mainaction} \mathcal{S}=\int d^4x\sqrt{-g}\left[
\frac{1}{2\kappa^2}F(R)-\frac{1}{2}\partial^{\mu}\phi\partial_{\mu}\phi-V(\phi)+\mathcal{L}_m
\right]\, ,
\end{equation}
where $\kappa^2=\frac{1}{8\pi G}=\frac{1}{M_p^2}$, with $G$ being
Newton's gravitational constant and $M_p$ stands for the reduced
Planck mass. Also, $\mathcal{L}_m$ denotes the Lagrangian density
of the perfect matter fluids that are present, which we will
assume that only radiation is present. The dark matter perfect
fluid will be composed solely by the axion particles present, with
the latter being identified with the scalar field $\phi$. Now,
with regard to the $F(R)$ gravity model, for phenomenological
reasons we will assume that it has the following form,
\begin{equation}\label{starobinsky}
F(R)=R+\frac{1}{M^2}R^2-\frac{\Lambda
\left(\frac{R}{m_s^2}\right)^{\delta}}{\zeta}\, ,
\end{equation}
with $m_s$ being defined as
$m_s^2=\frac{\kappa^2\rho_m^{(0)}}{3}$, also $\rho_m^{(0)}$ is the
energy density of cold dark matter at present day, and
$0<\delta<1$. Finally, $\zeta$ and $\gamma$ are some freely chosen
dimensionless constants for which we shall discuss at a latter
section. The $F(R)$ gravity model is composed by an $R^2$ model
which will control the primordial inflationary era, and by a power
law term $\sim R^{\delta}$, which eventually will control the
late-time dynamics. In fact the model (\ref{starobinsky}) can lead
to a viable dark energy era, as was shown in detail in
\cite{Oikonomou:2020qah,Odintsov:2020nwm,Oikonomou:2022bqb} so we
will not address the late-time dynamics issue here.

\begin{figure}
\centering
\includegraphics[width=18pc]{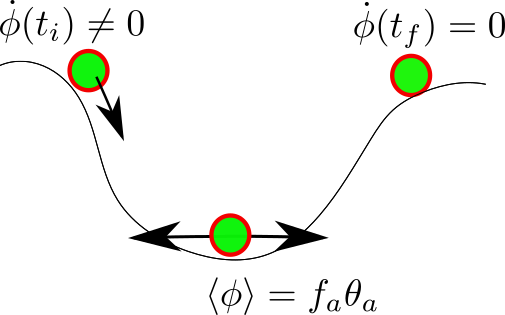}
\caption{The kinetic misalignment axion dynamics.}\label{plot2}
\end{figure}

With regard to the parameter $M$ appearing in the $R^2$ term, it
will be chosen to be $M= 1.5\times
10^{-5}\left(\frac{N}{50}\right)^{-1}M_p$, a value imposed by
inflationary phenomenological reasoning \cite{Appleby:2009uf},
with $N$ being the $e$-foldings number. Also the parameter
$\Lambda$ in Eq. (\ref{starobinsky}) is assumed to be of the same
order as the cosmological constant at present day. In this work we
shall consider a flat Friedmann-Robertson-Walker (FRW) background
with line element,
\begin{equation}
\label{metricfrw} ds^2 = - dt^2 + a(t)^2 \sum_{i=1,2,3}
\left(dx^i\right)^2\, ,
\end{equation}
so the field equations for the $f(R)$ gravity with the axion
scalar field in the presence of radiation are,
\begin{align}\label{eqnsofmkotion}
& 3 H^2F_R=\frac{RF_R-F}{2}-3H\dot{F}_R+\kappa^2\left(
\rho_r+\frac{1}{2}\dot{\phi}^2+V(\phi)\right)\, ,\\ \notag &
-2\dot{H}F=\kappa^2\dot{\phi}^2+\ddot{F}_R-H\dot{F}_R
+\frac{4\kappa^2}{3}\rho_r\, ,
\end{align}
\begin{equation}\label{scalareqnofmotion}
\ddot{\phi}+3H\dot{\phi}+V'(\phi)=0
\end{equation}
where $F_R=\frac{\partial F}{\partial R}$, while the ``dot''
denotes differentiation with respect to the cosmic time, and the
``prime'' denotes differentiation with respect to the axion scalar
field. With regard to the axion field, we shall consider the
misalignment axion \cite{Odintsov:2020nwm,Marsh:2015xka}, in which
case the axion should not be related to the QCD axion, but it is
some axion like particle, which we call axion. In the literature
there are two misalignment axion models, the canonical
misalignment model \cite{Marsh:2015xka} and the kinetic
misalignment model \cite{Co:2019jts,Co:2020dya,Barman:2021rdr}. In
this work we shall consider the effects of the kinetic
misalignment axion model on the inflationary dynamics of $F(R)$
gravity, and we shall see in which way it affects eventually the
duration of the inflationary era. In the kinetic misalignment
axion model, the axion has a primordial Peccei-Quinn $U(1)$
symmetry which is broken during the inflationary era. The fact
that the original Peccei-Quinn symmetry is broken is particularly
important for the inflationary phenomenology since no remnant
cosmic strings remain after inflation ends. This however was a
problem in standard QCD axion models, which is absent though in
all misalignment axion models. After the $U(1)$ symmetry is
broken, the axion acquires a large vacuum expectation value
$\langle \phi \rangle =\theta_a f_a$, where $\theta_a$ is the
initial misalignment angle, and $f_a$ is the axion decay constant.
The misalignment angle is in reality a dynamical field and can
take values in the range $0<\theta_a<1$, however, in the way that
it enters in the vacuum expectation value of the axion, it is not
considered as a dynamical field, but as an average value
throughout the whole Universe at the time of inflation. With
regard to the axion decay constant $f_a$, this parameter is of
fundamental phenomenological importance, and in conjunction with
the axion mass, constitute the two most important phenomenological
parameters for the axion dynamics. Regarding the axion having a
vacuum expectation value during inflation, this fact does not mean
that the axion is actually constant during inflation, but it
basically has small deviations about its vacuum expectation value,
different from the small oscillations about its vacuum expectation
value after the inflationary era ends. Let us describe in brief
the kinetic misalignment axion dynamics during inflation.
Schematically, this is depicted in Fig. \ref{plot2}.
As it can be seen in Fig. \ref{plot2}, the axion during inflation
has a small initial displacement from its vacuum expectation
value, and more importantly it has an non-zero initial velocity.
This initial velocity is what justifies the terminology
``kinetic''. The axion rolls down its potential and due to the
initial velocity it ends up uphill again, at a position different
than the one corresponding to the canonical misalignment axion
model, in which case the axion after it rolls down and reaches the
minimum, starts to oscillate around its vacuum expectation value.
Thus in the kinetic misalignment case, the axion ends up uphill
and it then rolls downhill until it reaches the minimum and
commences to oscillate around its vacuum expectation value when
the Hubble rate of the Universe becomes comparable to the axion
mass $H\sim m_a$. In the kinetic misalignment axion mechanism the
oscillations start at a later time compared to the canonical
misalignment mechanism, thus in the kinetic axion model, the
temperature at which oscillations commence is lower compared to
the canonical misalignment axion model. Let us briefly quantify
the above picture in terms of the potential and the axion mass.
Primordially, the axion potential has the following form,
\begin{equation}\label{axionpotentnialfull}
V(\phi )=m_a^2f_a^2\left(1-\cos (\frac{\phi}{f_a}) \right)\, ,
\end{equation}
and when the axion acquires a vacuum expectation value during
inflation, for small displacements from its vacuum expectation
value, its potential can be approximated as follows,
\begin{equation}\label{axionpotential}
V(\phi )\simeq \frac{1}{2}m_a^2\phi^2\, ,
\end{equation}
an approximation which is valid when $\phi\ll f_a$ or similarly
$\phi\ll \langle \phi \rangle $. Initially, when the axion is
uphill at both ends of the potential, we have $H\gg m_a$, however,
when the axion reaches the minimum for the second time, it starts
to oscillate around its vacuum expectation value when $H\leq m_a$,
and at that point the axion energy density redshifts as dark
matter \cite{Odintsov:2020nwm,Marsh:2015xka}. The most important
feature of the kinetic misalignment axion model is the fact that
initially the axion kinetic energy term $\dot{\phi}^2$ is quite
larger than the axion potential energy $\dot{\phi}^2\gg V$. This
continues until some time instance during the reheating, the
potential and the kinetic energy become comparable, when the
temperature of the Universe is of the order,
\begin{equation}\label{temperaturecondition}
\dot{\phi}(\tilde{T})=m_{a}(\tilde{T})\, ,
\end{equation}
at which temperature the axion has not a large kinetic energy
anymore so it becomes trapped in the potential barrier and the
oscillations around the minimum commence. The canonical
misalignment temperature when the oscillations start $T_*$ is
larger than $\tilde{T}$. So basically, when the axion mass is
larger than the Hubble rate $m_a(\tilde{T})\geq
H(\tilde{T})=\frac{3}{M_p}\sqrt{\frac{\pi^2}{10}}g_*\tilde{T}^2$,
the roll down and up of the axion occurs, and when
$m_a(\tilde{T})\leq
H(\tilde{T})=\frac{3}{M_p}\sqrt{\frac{\pi^2}{10}}g_*\tilde{T}^2$,
the axion starts to oscillate with abundance $\rho_a\sim
m_a(T=0)\frac{\dot{\phi}f_a^2}{s}$, where $s$ is the entropy
density, and $m_a(T=0)=m_a$ is the actual axion mass as a dark
matter particle. In general, in the kinetic axion misalignment
model, the axion dark matter mass is larger to the one compared to
the canonical misalignment axion case. A useful relation that
connects the axion mass with the axion decay constant is the
following,
\begin{equation}\label{axionmassfunctionofdecay}
m_a(T)=6\mathrm{meV}\frac{10^9\mathrm{GeV}}{f_a}\, ,
\end{equation}
and in order to the kinetic axion to account for the current dark
matter abundance, the axion decay constant must satisfy $f_a\leq
1.5\times10^{11}$GeV.

Let us further quantify the dynamics of the axion during inflation
during inflation, since this will be important for the study of
the inflationary phenomenology. Since initially, the kinetic
energy of the axion is quite larger than the potential energy,
that is, $\dot{\phi}^2\gg m_a^2\phi^2$, the field equation for the
axion becomes approximately,
\begin{equation}\label{eqnofmotionaxionkinetic}
\ddot{\phi}+3 H\dot{\phi}\simeq 0\, ,
\end{equation}
which can be solved to yield,
\begin{equation}\label{axionkineticfieldeqn}
\dot{\phi}\sim a^{-3}\, .
\end{equation}
So primordially, the energy density of the axion which is
$\rho_a=\frac{\dot{\phi}^2}{2}+V(\phi)\simeq
\frac{\dot{\phi}^2}{2}$ becomes $\rho_a\sim a^{-6}$. Thus this is
an era of kination for the axion dynamics, with its effective
equation of state parameter being $w=1$, which describes stiff
matter fluid. Hence the kinetic axion during inflation behaves as
a stiff perfect matter fluid. In the next section we shall
consider in a quantitative way the direct effects of the kinetic
misalignment axion scalar on the inflationary dynamics of $F(R)$
gravity and we shall reveal how the stiff axion fluid eventually
prolongs the inflationary era.

Before closing this section, let us note that the kinetic
misalignment axion mechanism is inherently related to the initial
explicit breaking of the Peccei-Quinn symmetry which is broken by
a higher dimensional effective operator in the same way as in the
Affleck-Dine mechanism.

\section{Dynamics of Inflation for the Kinetic Axion $F(R)$ Gravity Model}

Let us now use the results of the previous section in order to
determine the inflationary dynamics and the corresponding
phenomenology in terms of the slow-roll indices. Recall that in
the previous section we showed that the kinetic axion field
redshifts as a perfect matter fluid during inflation with a stiff
EoS, since $\rho_a\sim a^{-6}$, so its energy density is smaller
compared to the radiation fluid energy density. Assuming a low
scale inflationary era, with the Hubble rate during inflation
being of the order $H_I=10^{13}$GeV, let us investigate which
terms effectively dominate in the field equations of the kinetic
axion $F(R)$ gravity. The Ricci scalar takes quite large values
for $H_I=10^{13}$GeV, thus the $R^2$ dominates the evolution
roughly speaking. Let us see this in some detail, and recall that
$m_s^2\simeq 1.87101\times 10^{-67}$eV$^2$ and also the parameter
$M$ which appears in the $R^2$ term in Eq. (\ref{starobinsky}) is
$M= 1.5\times 10^{-5}\left(\frac{N}{50}\right)^{-1}M_p$
\cite{Appleby:2009uf}, hence by roughly taking for $N\sim 60$, $M$
is of the order $M\simeq 3.04375\times 10^{22}$eV. Also, due to
the fact that during inflation the slow-roll conditions are
satisfied, we approximately have $R\simeq 12 H^2$ and therefore
$R\sim 1.2\times 10^{45}$eV$^2$. Furthermore, $M_p\simeq 2.435
\times 10^{27}$eV, and also the parameter $\Lambda$ is of the
order of the cosmological constant at present day, that is,
$\Lambda\simeq 11.895\times 10^{-67}$eV$^2$. Finally, the vacuum
expectation value of the axion is roughly of the same order as the
axion decay constant, therefore $\langle \phi \rangle
=\phi_i\simeq \mathcal{O}(10^{15})$GeV and approximately
$m_a\simeq \mathcal{O}(10^{-14})$eV. Thus,  the potential term is
of the order $\kappa^2V(\phi_i)\sim \mathcal{O}(8.41897\times
10^{-36})$eV$^{2}$, while the two curvature terms $R$ and $R^2$
are of the order, $R\sim 1.2\times \mathcal{O}(10^{45})$eV$^2$ and
also $R^2/M^2\sim \mathcal{O}(1.55\times 10^{45})$eV$^2$ and the
power-law curvature term is of the order $\frac{\Lambda
\left(\frac{R}{m_s}\right)^{0.1}}{0.2}\sim
 \mathcal{O}(10^{-55})$eV$^2$ for $\delta=0.1$ and $\zeta=0.2$,
with the latter being phenomenologically acceptable values. Also,
during inflation the radiation density term $\kappa^2\rho_r \sim
\kappa^2 e^{-4N}$ and also a similar relation applies for the
kinetic misalignment axion energy density $\rho_a$ and the
corresponding term scales as $\kappa^2\rho_r \sim \kappa^2
e^{-6N}$. Therefore, at the equations of motion level, the
resulting theory is basically effectively described by a vacuum
$R^2$, in which case,
\begin{equation}\label{effectivelagrangian2}
F(R)\simeq R+\frac{1}{M^2}R^2\, .
\end{equation}
Then the field equations take the form
\begin{equation}\label{patsunappendix}
\ddot{H}-\frac{\dot{H}^2}{2H}+\frac{H\,M^2}{2}=-3H\dot{H}\, .
\end{equation}
and due to the slow-roll conditions,
\begin{equation}\label{patsunappendix1}
-\frac{M^2}{6}=\dot{H}\, ,
\end{equation}
which has as a solution,
\begin{equation}\label{quasidesitter}
H(t)=H_I-\frac{M^2}{6} t\, ,
\end{equation}
which is a quasi-de Sitter solution, with $H_I$ being an arbitrary
integration constant, with profound physical significance since
this is the scale of inflation.

Now one might consider that effectively the kinetic axion does not
affect at all the dynamics of inflation, however this is not true.
It is certain that the axion does not control the Hubble rate at
the level of the equations of motion for sure, however the
dynamics of inflation are not affected only by the background
evolution. As we will show the axion may affect inflation in two
ways, firstly it may directly affect the scalar curvature
perturbations and secondly it prolongs the inflationary era, since
the axion effective equation of state is $w=1$ so inflation is
prolonged as we show shortly.

The cosmological scalar curvature perturbations are dynamically
quantified by the slow-roll indices, which for the $f(R,\phi)$
theory at hand are defined to be
\cite{Hwang:2005hb,reviews1,Odintsov:2020thl},
\begin{equation}
\label{restofparametersfr}\epsilon_1=-\frac{\dot{H}}{H^2}, \quad
\epsilon_2=\frac{\ddot{\phi}}{H\dot{\phi}}\, ,\quad \epsilon_3=
\frac{\dot{F}_R}{2HF_R}\, ,\quad
\epsilon_4=\frac{\dot{E}}{2H\,E}\,
 ,
\end{equation}
where the function $E$ for the $f(R,\phi)$ theory at hand has the
following form,
\begin{equation}\label{eparameter}
E=F_R+\frac{3\dot{F}_R^2}{2\kappa^2\dot{\phi}^2}\, .
\end{equation}
Now the most important effect that the kinetic axion theory brings
along in the $F(R)$ gravity is contained in the parameter
$\epsilon_2$. Since the axion obeys the stiff scalar differential
equation (\ref{eqnofmotionaxionkinetic}), this means that in our
case, the slow-roll parameter $\epsilon_2$ takes the value
$\epsilon_2=-3$, therefore the axion obeys a constant-roll
condition in its dynamics. The question is, does $\epsilon_2$
affect the inflationary dynamics? As we now show, at leading
order, the contribution of the axion field is elegantly cancelled
in the observational indices of inflation and specifically when
the spectral index of the primordial scalar curvature
perturbations. To this end, let us present the details of the
calculation of the parameter $\epsilon_4$ in which the dynamics of
the axion is found. We have explicitly at leading order during
inflation,
\begin{equation}\label{epsilon41}
E\simeq \frac{3\dot{F}_R^2}{2\kappa^2\dot{\phi}^2}\, ,
\end{equation}
so $\epsilon_4$ is approximately equal to,
\begin{equation}\label{epsilon42}
\epsilon_4\simeq
\frac{3}{2\kappa^2}\frac{2\dot{F}_R\ddot{F}_R\dot{\phi}^2-\dot{F}_R^2\dot{\phi}\ddot{\phi}}{\dot{\phi}^4}\,
,
\end{equation}
which is simplified to,
\begin{equation}\label{epsilon43}
\epsilon_4\simeq
\frac{\ddot{F}_{RR}}{H\dot{F}_R}-\frac{\ddot{\phi}}{H\dot{\phi}}=\frac{\ddot{F}_{RR}}{H\dot{F}_R}-\epsilon_2\,
.
\end{equation}
Let us further elaborate on the parameter $\epsilon_4$ which after
some algebra is written as follows,
\begin{equation}\label{epsilon4final}
\epsilon_4\simeq -\frac{24
F_{RRR}H^2}{F_{RR}}\epsilon_1-3\epsilon_1+\frac{\dot{\epsilon}_1}{H\epsilon_1}-\epsilon_2\,
.
\end{equation}
The term $\dot{\epsilon}_1$ can be written as,
\begin{equation}\label{epsilon1newfiles}
\dot{\epsilon}_1=-\frac{\ddot{H}H^2-2\dot{H}^2H}{H^4}=-\frac{\ddot{H}}{H^2}+\frac{2\dot{H}^2}{H^3}\simeq
2H \epsilon_1^2\, ,
\end{equation}
hence $\epsilon_4$ becomes,
\begin{equation}\label{finalapproxepsilon4}
\epsilon_4\simeq -\frac{24
F_{RRR}H^2}{F_{RR}}\epsilon_1-\epsilon_1-\epsilon_2\, .
\end{equation}
Upon introducing $x$ we have,
\begin{equation}\label{parameterx}
x=\frac{48 F_{RRR}H^2}{F_{RR}}\, ,
\end{equation}
and $\epsilon_4$ can be written in terms of it as follows,
\begin{equation}\label{epsilon4finalnew}
\epsilon_4\simeq -\frac{x}{2}\epsilon_1-\epsilon_1-\epsilon_2\, .
\end{equation}
Now for the $f(R,\phi)$ gravity, the scalar spectral index of the
scalar curvature perturbations is
\cite{Hwang:2005hb,reviews1,Odintsov:2020thl},
\begin{equation}\label{scalarspectralindex}
n_{\mathcal{S}}=1-4\epsilon_1-2\epsilon_2+2\epsilon_3-2\epsilon_4\,
,
\end{equation}
thus by substituting the expression for $\epsilon_4$ we obtained
in Eq. (\ref{finalapproxepsilon4}), we can see that elegantly the
contribution of $\epsilon_2$ cancels, thus the spectral index
takes the form,
\begin{equation}\label{spectralindexfinalform}
n_{\mathcal{S}}\simeq 1-(2-x)\epsilon_1+2\epsilon_3\, .
\end{equation}
Also the scalar-to-tensor ratio for the case at hand is equal to
\cite{Hwang:2005hb,reviews1,Odintsov:2020thl},
\begin{equation}\label{tensortoscaalrratio}
r\simeq 48\epsilon_1^2\, .
\end{equation}
Since the dominant part of the $F(R)$ gravity during inflation is
an $R^2$ gravity, the term $x$ is equal to zero thus, the scalar
spectral index is greatly simplified. For the quasi-de Sitter
solution at hand, the first slow-roll index is easily calculated
to be,
\begin{equation}\label{epsilon1indexanalytic}
\epsilon_1=-\frac{6 M^2}{\left(M^2 t-6 H_I\right)^2}\, ,
\end{equation}
and by solving the algebraic equation $\epsilon_1(t_f)=1$, the
time instance at which inflation ends is,
\begin{equation}\label{finaltimeinstance}
t_f=(6 H_I + \sqrt{6} M)/M^2\, .
\end{equation}
Using the definition of the $e$-foldings number $N$,
\begin{equation}\label{efoldingsnumber}
N=\int_{t_i}^{t_f}H(t)dt\, ,
\end{equation}
the time instance at which inflation commences is,
\begin{equation}\label{ti}
t_i=\frac{2 \sqrt{9 H_I^2-3 M^2 Y}+6 H_I}{M^2}\, ,
\end{equation}
so the first slow-roll index at first horizon crossing is,
\begin{equation}\label{epsilon1lambdaind}
\epsilon_1(t_i)=\frac{1}{1+2N}\, ,
\end{equation}
hence at leading order in terms of the $e$-foldings number, the
spectral index and the tensor-to-scalar ratio take the form
$n_s\sim 1-\frac{2}{N}$ and $r\sim \frac{12}{N^2}$. Now for $N\sim
60$ the resulting phenomenology is identical to the Starobinsky
model, however the axion stiff equation of state causes another
effect on inflation. Basically it prolongs the inflationary era to
some extent as we now evince.
\begin{figure}
\centering
\includegraphics[width=18pc]{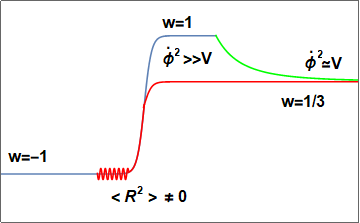}
\caption{The kinetic misalignment axion $F(R)$ gravity total EoS
dynamical evolution. The cosmological system reaches an unstable
de Sitter point in both the vacuum $R^2$ gravity and kinetic
misalignment axion $F(R)$ gravity. Eventually, the $\langle R^2
\rangle $ fluctuations make the system to be repelled from the de
Sitter attractor, and the cosmological system enters the reheating
era controlled by the $\langle R^2 \rangle $ fluctuations. In the
presence of the kinetic axion, after the $\langle R^2 \rangle $
fluctuations cause the system to be repelled from the de Sitter
attractor, the cosmological system does not enter the reheating
era directly, but the kinetic term dominates the evolution and the
background EoS is not the one corresponding to an ordinary
reheating era $w=1/3$ but it corresponds to a stiff era with
$w=1$. The system stays in this stiff era and the ordinary
reheating era commences when the axion oscillations
begin.}\label{kinetic}
\end{figure}
As inflation comes to an end near the time instance $t_f$, the
background total EoS of the Universe is no longer described by a
quasi-de Sitter EoS hence the stiff EoS of the axion describes the
Universe, since the matter perfect fluids become more dominant
slowly-by-slowly. Therefore the total EoS parameter of the
background evolution approaches the stiff EoS value $w=1$. This
fact prolongs the inflationary era, causing the $e$-foldings
number to be larger than 60. The physical picture behind the
increase of the $e$-foldings number relies on the combined
presence of the $R^2$ term and the large kinetic term of the
kinetic misalignment axion. In standard $R^2$ gravity, inflation
tends to its end when the curvature fluctuations $\langle R^2
\rangle $ become quite strong and make the de Sitter attractor
unstable. This phenomenological picture is possible only to the
$R^2$ gravity, and as it was shown in Ref.
\cite{Odintsov:2017tbc}, the Starobinsky model has an unstable de
Sitter attractor. Let us show this in brief, see also
\cite{Odintsov:2017tbc} for more details. By introducing the
dimensionless variables,
\begin{equation}\label{variablesslowdown}
x_1=-\frac{\dot{F_R}(R)}{F_R(R)H},\,\,\,x_2=-\frac{F(R)}{6F(R)H^2},\,\,\,x_3=
\frac{R}{6H^2}\, ,
\end{equation}
and by using the $e$-foldings number as a dynamical variable
instead of the cosmic time, the field equations of vacuum $F(R)$
gravity can be written in terms of the following dynamical system,
\begin{align}\label{dynamicalsystemmain}
& \frac{\mathrm{d}x_1}{\mathrm{d}N}=-4-3x_1+2x_3-x_1x_3+x_1^2\, ,
\\ \notag &
\frac{\mathrm{d}x_2}{\mathrm{d}N}=8+m-4x_3+x_2x_1-2x_2x_3+4x_2 \, ,\\
\notag & \frac{\mathrm{d}x_3}{\mathrm{d}N}=-8-m+8x_3-2x_3^2 \, ,
\end{align}
with the dynamical parameter $m$ being equal to,
\begin{equation}\label{parameterm}
m=-\frac{\ddot{H}}{H^3}\, .
\end{equation}
The dynamical system (\ref{dynamicalsystemmain}) is autonomous
when the parameter $m$ takes constant values, and for a quasi-de
Sitter evolution $a(t)=e^{H_0 t-H_i t^2}$ the parameter $m$ is
equal to zero. The total EoS of the cosmological system is defined
as \cite{reviews1}
\begin{equation}\label{weffoneeqn}
w_{eff}=-1-\frac{2\dot{H}}{3H^2}\, ,
\end{equation}
and it can directly be expressed in terms of the variable $x_3$ in
the following way,
\begin{equation}\label{eos1}
w_{eff}=-\frac{1}{3} (2 x_3-1)\, .
\end{equation}
Now, by performing a fixed point analysis of the dynamical system
for $m=0$, we easily obtain the fixed points,
\begin{equation}\label{fixedpointdesitter}
\phi_*^1=(-1,0,2),\,\,\,\phi_*^2=(0,-1,2)\, ,
\end{equation}
and the corresponding eigenvalues of the matrix which corresponds
to the dynamical system for $\phi_*^1$ are $(-1, -1, 0)$, while in
the case of the fixed point $\phi_*^2$ these are $(1, 0, 0)$.
Therefore, the dynamical system possesses two non-hyperbolic, but
the fixed point $\phi_*^1$ is stable and in contrast, the fixed
point $\phi_*^2$ is unstable, with the latter being the most
interesting fixed point from a phenomenological point of view. It
is noticeable that for both the fixed points, we have $x_3=2$
hence, from Eq. (\ref{eos1}), we get $w_{eff}=-1$. This feature
basically shows that both fixed points are de Sitter fixed points.
As we already mentioned, the second de Sitter fixed point, namely,
$\phi_*^2=(0,-1,2)$, is the most interesting phenomenologically,
since for this equilibrium, the conditions $x_1\simeq 0$ and
$x_2\simeq -1$ yield,
\begin{align}\label{caseidiffseqns1}
-\frac{\mathrm{d}^2F}{\mathrm{d}R^2}\frac{\dot{R}}{H\frac{\mathrm{d}F}{\mathrm{d}R}}\simeq
0,\,\,\,-\frac{F}{H^2\frac{\mathrm{d}F}{\mathrm{d}R}6}\simeq -1\,
.
\end{align}
Using the slow-roll approximation during inflation for the Ricci
scalar curvature $R\simeq 12 H^2$, for the quasi-de Sitter
evolution, we can write the second differential equation as
follows,
\begin{equation}\label{seconddiff}
F\simeq \frac{\mathrm{d}F}{\mathrm{d}R} \frac{R}{2}\, ,
\end{equation}
which when solved it yields,
\begin{equation}\label{approximatersquare}
F(R)\simeq \alpha R^2\, ,
\end{equation}
where $\alpha$ is some arbitrary integration constant, which
describes which $F(R)$ gravity generates the quasi-de Sitter
evolution. Clearly the $R^2$ model possesses an unstable de Sitter
point. Thus when this unstable de Sitter attractor is reached, the
system is repelled from it in the phase space. The time instance
for which this happens is determined roughly by the condition
$\epsilon_1(t_f)=1$. Now in the presence of the large axion
kinetic term which dominates over its potentials, things are
somewhat different when the cosmological system reaches the de
Sitter attractor. Particularly, the cosmological system initially
is controlled by the $R^2$ term so it reaches the quasi-de Sitter
attractor. However, when it is repelled from the unstable de
Sitter attractor point, the cosmological system does not enter
directly the reheating era and the $\langle R \rangle $ reheating
fluctuations do not commence directly, but the kinetic term which
was subdominant, dominates over the $R^2 $ term and thus control
the dynamics at the end of inflation, after the cosmological
system is repelled from the quasi-de Sitter attractor. Thus the
end of inflation is somewhat prolonged for the kinetic axion $R^2$
model. This can be schematically seen in Fig. \ref{kinetic}, in
which it is shown that in the ordinary $R^2$ model, the system
after it reaches the unstable de Sitter attractor, the $\langle
R^2 \rangle $ fluctuations make the system to be repelled from the
attractor, and the cosmological system enters the reheating era
controlled by the $\langle R \rangle $ fluctuations. In the
presence of the kinetic axion, after the $\langle R^2 \rangle $
fluctuations cause the system to be repelled from the de Sitter
attractor, the cosmological system does not enter directly the
reheating era, but the kinetic term dominates the evolution and
the background EoS is not the one corresponding to an ordinary
reheating era $w=1/3$ but it corresponds to a stiff era with
$w=1$. The system stays in this stiff era and the ordinary
reheating era commences when the axion oscillations begin, so when
$\dot{\phi}^2\sim V$, at which point the axion redshifts as
ordinary dark matter and the radiation fluid controls the
evolution thereafter. Thus the number of $e$-foldings is somewhat
extended in the kinetic axion $F(R)$ gravity picture.

Another striking feature of the kinetic axion $F(R)$ gravity model
is the fact that the $R^2$ term actually enhances significantly
the kinetic axion physics, delaying further the kinetic axion to
start its oscillations. In a near future work we shall demonstrate
using a dynamical systems approach how this can happen.
\begin{figure}
\centering
\includegraphics[width=26pc]{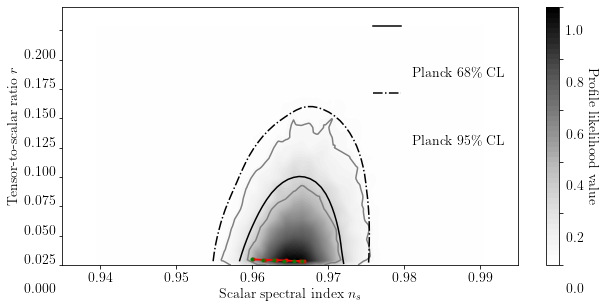}
\caption{The Planck likelihood curves and the kinetic axion $F(R)$
gravity model (red dots) and the vacuum $R^2$ model (green dots).
The kinetic $R^2$ model serves as a viable deformation of the
vacuum Starobinsky model.} \label{planck}
\end{figure}
Now let us quantify the qualitative picture we described above,
and let us see how the stiff era affects the $e$-foldings number,
thus somewhat extending inflation for some $e$-foldings. As we
will show, this feature is strongly affected by the reheating
temperature. In a general setting, the $e$-foldings number for a
primordial scalar mode with wavenumber $k$, which became
superhorizon at the beginning of inflation, is equal to
\cite{Adshead:2010mc},
\begin{equation}\label{generalefoldingsnumber}
\frac{a_kH_k}{a_0H_0}=e^{-N}\frac{H_ka_{end}}{a_{reh}H_{reh}}\frac{H_{reh}a_{reh}}{a_{eq}H_{eq}}\frac{H_{eq}a_{eq}}{a_{0}H_{0}}\,
,
\end{equation}
with $a_k$ and $H_k$ being the scale factor and the Hubble rate at
the time instance where the primordial mode $k$ became
superhorizon at the beginning of inflation (at first horizon
crossing), $a_{end}$ stands for the scale factor at the end of the
inflationary era, and finally $a_{reh}$ and $H_{reh}$ denote the
scale factor and the Hubble rate when the reheating era ends.
Furthermore, $a_{eq}$ and $H_{eq}$ denote the scale factor and the
Hubble rate at the time instance that the matter-radiation
equality occurs, and moreover $a_0$ and $H_0$ denote the present
day scale factor and the Hubble rate respectively. Now, if near
the end of inflation, the total EoS parameter is $w$ (different
from the value $w=1/3$), we get,
\begin{equation}\label{aux1}
\ln \left(\frac{a_{end}H_{end}}{a_{reh}H_{reh}}
\right)=-\frac{1+3w}{6(1+w)}\ln
\left(\frac{\rho_{reh}}{\rho_{end}} \right)\, ,
\end{equation}
with $H_{end}$ being the Hubble rate when inflation ends, and the
energy densities $\rho_{end}$ and $\rho_{reh}$ stand for the total
energy density of the Universe when inflation ends and when the
reheating era ends. Note that for the derivation of Eq.
(\ref{aux1}), we assumed that the total EoS parameter at the end
of the reheating era and at the end of inflation, is constant and
equal to $w$. Then, when the $\langle R^2 \rangle$ commence,
causing instability to the de Sitter period, the constant EoS
stiff era of the kinetic axion commences, so the $e$-foldings
number of the inflationary era is extended as follows
\cite{Adshead:2010mc},
\begin{equation}\label{efoldingsmainrelation}
N=56.12-\ln \left( \frac{k}{k_*}\right)+\frac{1}{3(1+w)}\ln \left(
\frac{2}{3}\right)+\ln \left(
\frac{\rho_k^{1/4}}{\rho_{end}^{1/4}}\right)+\frac{1-3w}{3(1+w)}\ln
\left( \frac{\rho_{reh}^{1/4}}{\rho_{end}^{1/4}}\right)+\ln \left(
\frac{\rho_k^{1/4}}{10^{16}\mathrm{GeV}}\right)\, ,
\end{equation}
with $\rho_k$ being the Universe's total energy density at the
beginning of the inflationary era, exactly when the mode $k$
became superhorizon. We shall also assume that the pivot scale
$k_*$ is $k_*=0.05$Mpc$^{-1}$ and furthermore, we shall assume
that the degrees of freedom of particles $g_*$ during the
inflationary era, just after this era is nearly constant. Thus the
energy density of the Universe at a temperature $T$ is equal to
$\rho=\frac{\pi^2}{30}g_*T^4$. Hence, the expression of Eq.
(\ref{efoldingsmainrelation}) can be rewritten in terms of the
temperatures at the various epochs and not in terms of the energy
densities.
\begin{table}[h!]
  \begin{center}
    \label{table1}
    \begin{tabular}{|r|r|r|}
     \hline
      \textbf{$e$-foldings number and Inflationary Indices} & \textbf{$T_R=10^{12}$GeV} & \textbf{$T_R=10^{7}$GeV} \\
           \hline
           $e$-foldings number $N$ & 65.3439 & 61.5063\\ \hline
           Spectral index $n_{\mathcal{S}}$ & 0.969393 & 0.967483 \\ \hline
      Tensor-to-Scalar Ratio $r$ & 0.00281042 & 0.00317206 \\ \hline
      \end{tabular}
  \end{center}
      \caption{\emph{\textbf{The $e$-foldings number for the kinetic axion $F(R)$ gravity model for various reheating temperatures, to be compared with the standard
    $R^2$ model results $n_{\mathcal{S}}=0.966667$ and $r=0.00333333$ and a standard reheating scenario.}}}
\end{table}
In effect if the total number of $e$-foldings changes, the
parameter $M$ coupled to the $R^2$ gravity will also be somewhat
affected and this should be taken into account for the
inflationary phenomenology of the current model. In Table
\ref{table1} we present the  phenomenological behavior of the
basically prolonged $R^2$ inflationary model, for three reheating
temperatures, namely a large reheating temperature
$T_R=10^{12}$GeV, and an intermediate reheating temperature
$T_R=10^{7}$GeV. The perspective of having low reheating
temperatures is already discussed in the literature, even having
MeV scale reheating temperatures, see for example
\cite{Hasegawa:2019jsa}. As it can be seen, in all cases, the
inflationary era is prolonged and the results are different from
the standard $R^2$ model for $N=60$ with the changes being of the
order 15$\%$ for the case of the tensor-to-scalar ratio. Also as
expected, since the inflationary era generated by the kinetic
axion $F(R)$ gravity theory is a deformation of the $R^2$ model,
it produces a viable phenomenology. This can be seen in Fig.
\ref{planck} where we confront the kinetic axion $F(R)$ gravity
model with the Planck likelihood curves for various reheating
temperatures in the range $10^{7}-10^{12}$GeV. As it can be seen,
the model is well fitted in the sweet spot of the Planck data. In
the plots, the green dots correspond to the vacuum $R^2$ model,
and the red dots correspond to the kinetic axion $R^2$ model. As
it can be seen, the kinetic axion $R^2$ model is a measurable
deformation of the vacuum $R^2$ model.

\section{Conclusions}

In this work we investigated how a kinetic misalignment axion can
affect the inflationary era generated by an $R^2$ model of $F(R)$
gravity. In the context of the kinetic misalignment axion, the
primordial $U(1)$ Peccei-Quinn symmetry is broken in the axion
sector during inflation, thus the axion has a non-zero vacuum
expectation value, however it also possesses a large kinetic
energy. The kinetic energy term of the axion dominates over its
potential, however during inflation and at the equations of motion
level, the vacuum $R^2$ model dominates the evolution. Thus the
axion may affect the dynamics of the inflationary era at the
cosmological perturbations level, through the second slow-roll
index. Due to the dominance of the axion's kinetic energy over its
potential, the axion evolves in a constant-roll way, thus the
second slow-roll index is constant and large. We calculated the
observational indices including the kinetic axion effects, and as
we showed, the contribution of the second slow-roll index
elegantly cancels. Thus, at the cosmological perturbations level,
the kinetic axion does not affect the $R^2$ inflationary era.
However, the kinetic axion affects the duration of the
inflationary era, causing in some cases 15$\%$ differences in the
tensor-to-scalar ratio compared with the vacuum $R^2$ model. This
change is due to the fact that the kinetic axion has an EoS
parameter which corresponds to that of a stiff era. As the $R^2$
inflationary era reaches its unstable quasi-de Sitter attractor in
the phase space, the kinetic axion starts to dominate the
evolution over the $R^2$ term, thus the Universe enters a stiff
evolution era, and era of kination with background total EoS
parameter $w=1$. This stiff background directly affects the
$e$-foldings number, thus extending the inflationary era up to $5$
$e$-foldings in some cases, and quantitatively in some cases this
amounts to a decrease of the tensor-to-scalar ratio of about
15$\%$ compared to the vacuum $R^2$ model. A particularly
interesting extension of this work is to further consider in the
Lagrangian an Einstein-Gauss-Bonnet term. Due to the fact that the
axion is not constant during inflation, but it is fluctuating
around its vacuum expectation value, the Einstein-Gauss-Bonnet
does not trivially vanish, thus it would be interesting to
investigate the consequences of the kinetic axion in this class of
theories. In fact, it would be furthermore interesting to
investigate the late-time evolution of the unified model, because
when the axion starts to oscillate around its vacuum expectation
value, it redshifts as dark matter. These issues shall be
addressed in a future work.

\end{document}